\documentclass[aps,prl,reprint,superscriptaddress]{revtex4-2}

\usepackage{amsmath}
\usepackage{amssymb}
\usepackage{bm}
\usepackage{graphicx}

\begin{document}

\title{Occupancy Driven Phase Transitions in Gapped Systems}

\author{Ian Madden}
\affiliation{Department of Electrical Engineering and Computer Science, University of Missouri, Columbia,     65211 MO.}
\author{Alex Kociubinski}
\affiliation{Department of Electrical Engineering and Computer Science, University of Missouri, Columbia,     65211 MO.}
\author{Wouter Montfrooij} 
\affiliation{Department of Physics and Astronomy, University of Missouri, Columbia, 65211 MO.}
\date{\today}

\begin{abstract}
Systems characterized by an energy gap between the ground state and excited states support a growing number of thermal excitations upon raising the temperature. Provided these excitations interact with each other, their number affects the size of the energy gap akin to a harmonic oscillator where the oscillation frequency is damping dependent. Here we show that once a critical number of excitations is reached, the excitation population becomes thermodynamically unstable and the system transitions between two states. This transition is marked by a divergence in the specific heat, which does not have to be accompanied by an order parameter, nor by the full closure of the gap. We illustrate the thermodynamics using a simple two level system and superfluid  $^4$He as examples, and we argue that the hidden order transition in URu$_2$Si$_2$ is an example of such an occupancy driven transition in the absence of an order parameter.
\end{abstract}
\maketitle
\begin{figure}[t]
	\includegraphics*[viewport=80 90 555 345,width=85mm,clip]{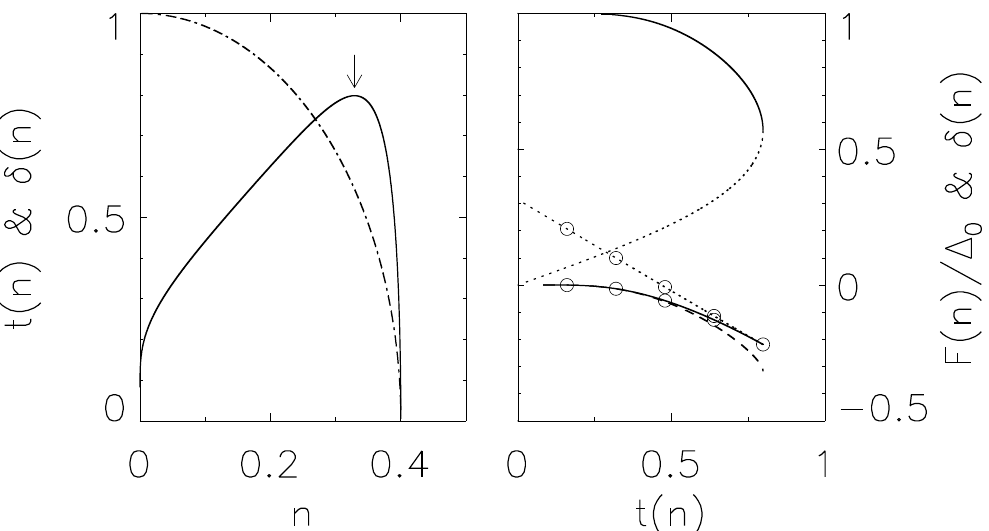}
	\caption{Left panel: the reduced temperature (solid line) and gap (dashed line) versus occupancy. The arrow points to the maximum of $t(n)$. Right: The free energy [eqn (\ref{freeenergy})] displays a minimum (solid curve with symbols) and a maximum (dotted) for each temperature. The dashed curve is the incorrect free energy curve based on $E(n)=n\Delta(n)$. Also shown is the gap as a function of $T$ (solid line: free energy minimum; dotted curve: maximum). $\alpha$= 2.5 was used for this figure.}
	\label{twolevel}
\end{figure}
\begin{figure}[t]
	\includegraphics*[viewport=120 45 570 470,width=85mm,clip]{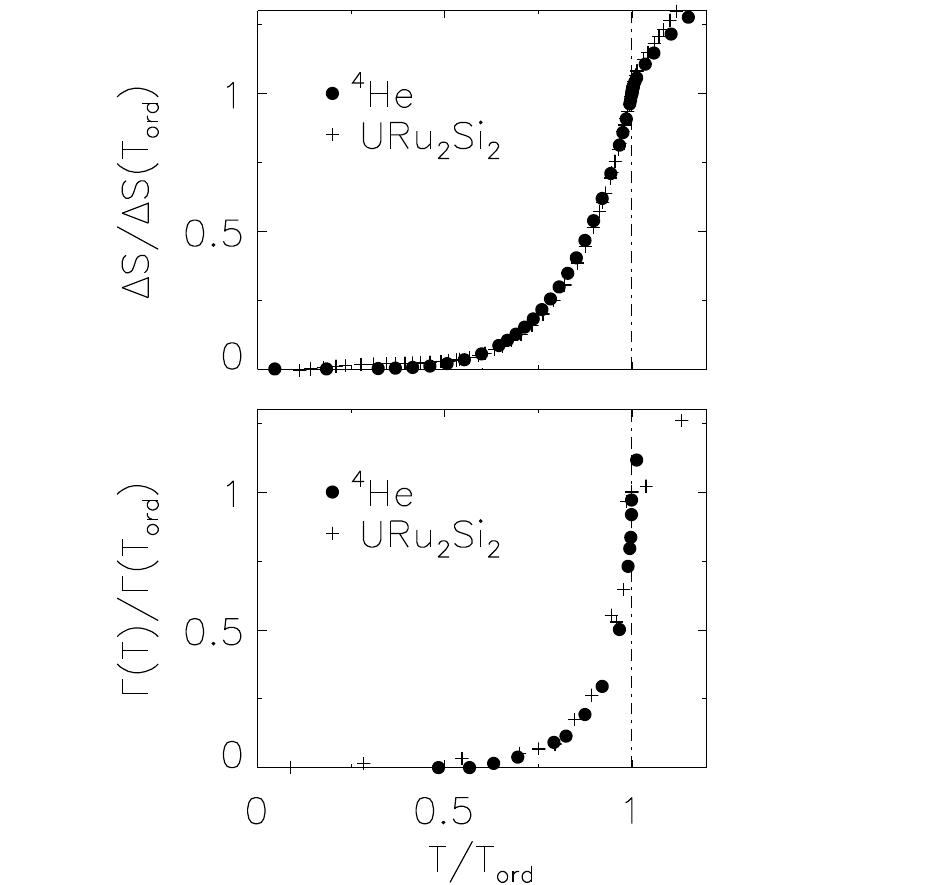}
	\caption{Figure adapted from Ref. \cite{Montfrooij2024}. Top panel: Entropy versus temperature for URu$_2$Si$_2$ [$\Delta$S= S-S(T=1 K)] and $^4$He ($\Delta$S= S), made to coincide at the ordering temperature. Bottom panel: The (scaled) damping rates of the roton minimum and the incommensurate gap minimum \cite{Broholm1991} in  URu$_2$Si$_2$. The temperature evolution of both systems is essentially identical.}
	\label{damping}
\end{figure}
\begin{figure}[t]
	\includegraphics*[viewport=75 85 400 315,width=65mm,clip]{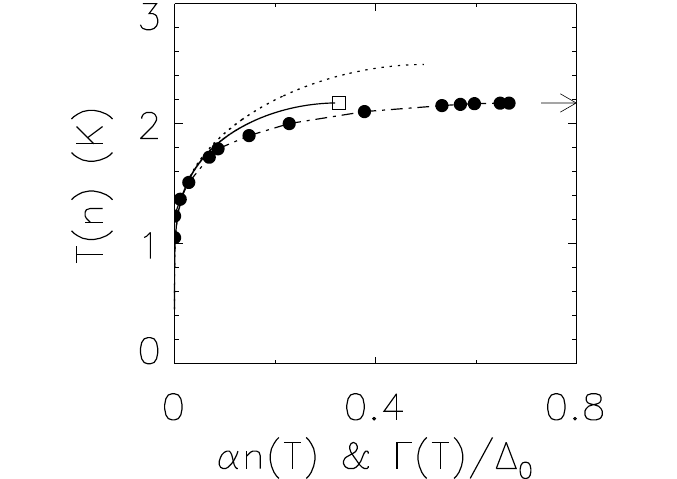}
	\caption{$T(n)$ for superfluid helium versus $\Gamma(T)=\Delta_0 \alpha n_{\text{rot}}(T)$ using the analytically determined number of roton excitations from eqn (\ref{nrot}) as input (dotted) and for  eqn (\ref{nrot})  including the energy shift $\delta \epsilon=\Delta_0 \beta n_{\text{rot}}(T)$ (solid curve). The measured $\Gamma(T)$ \cite{Montfrooij1997} are displayed as filled symbols, while the arrow indicates $T_{\lambda}$. Note the similar approach of all three curves, representing the termination of the self-consistent solution (calculations) and the process of the actual roton excitation becoming an ordinary density fluctuation. The square indicates the point where the roton density equals the fluid density.}
	\label{lambda}
\end{figure}

When the energy of an excitation is affected by the presence of other excitations, a point can be reached where the population of excitations becomes unstable against their number. Bloch showed \cite{Bloch1962,Bloch1963}, by retaining higher-order terms in the spin Hamiltonian that the magnon population in gapless magnets becomes unstable at a temperature close to the transition temperature.  J\"ackle demonstrated \cite{Jackle1972} that a gapped system (superfluid helium) can exhibit a similar breakdown when excitations can form bound pairs, although the required binding energy was substantially larger than experimentally observed. Here we demonstrate that occupancy-driven instabilities are a generic consequence of interactions between excitations and do not require bound-state formation. In particular, we show that damping arising from Landau--Khalatnikov-type interactions can itself drive such an instability, which we identify as an fold instability where the free energy minimum and maximum solutions merge. We establish this mechanism in a minimal two-level system and illustrate its implications for the superfluid transition in $^4$He and the hidden-order transition in URu$_2$Si$_2$.\\
Consider a two-level system characterized by a temperature dependent energy gap $\Delta(T)$. The thermal occupancy $n(T)$ of the upper level is given by
\begin{equation}
n(T)=\frac{1}{1+\exp[\Delta(T)/(k_BT)]},
\label{occ}
\end{equation}
where $k_B$ is Boltzmann's constant. When the gap is absent, or when $T$ becomes very large, $n(T)$= 1/2 and both levels are equally likely to be occupied. When the temperature dependence of the gap is a function of the occupancy (as is the case for helium), the gap becomes a function of occupancy: $\Delta(n)$. We invert eqn (\ref{occ}) to obtain
\begin{equation}
t(n)=\frac{\delta(n)}{\ln[(1-n)/n]},
\label{invert}
\end{equation}
where we have introduced the dimensionless quantities $\delta=\Delta(n)/\Delta_0$ and $t= k_BT/\Delta_0$ with $\Delta_0$ the gap at T= 0. Based on the experimental results of $^4$He where a damped harmonic oscillator dependence of the energy gap was observed (to be discussed), we choose a gap of the form  $\delta(n)=\sqrt{[1+\beta n(T)]^2-[\Gamma(T)/\Delta_0]^2}$ with $\Gamma(T) =\alpha n(T)$ and $\alpha$ and $\beta$ constants determined by the details of the interaction \cite{rates}.  The energy shift $\beta n(T)$ incorporates the Hartree term ($\sim g_4$) as well as the energy shift due to scattering by thermal excitations  ($\sim |g_4|^2$) . This thermal scattering also gives rise to the $\Gamma(T)$ term. The energy shift
was studied by Ruvals \cite{Ruvalds1971} and by J\"ackle \cite{Jackle1972} for $^4$He who showed that (with $\alpha= 0$) this shift leads to an instability in the gas of roton excitations and suggested this instability was associated with $T_{\lambda}$. However, the $\beta$ value required ($\beta <$ -2) is not consistent with the experimental value of $\beta \approx$ -0.8. We show that the $\Gamma$-term is also required to drive the system to an instability. To do so, we first set $\beta$=0 and study
\begin{equation}
t(n)=\frac{\sqrt{1-\alpha^2 n^2}}{\ln[(1-n)/n]}.
\label{tn}
\end{equation}
The above equation yields different behavior depending on whether $\alpha$ is less than, equal to, or greater than 2: $\alpha$=2 would see the gap close when the occupancy reaches 1/2 and $T= \infty$. For $\alpha > 2$, the gap closes at $n=1/\alpha$. However, the system becomes unstable before this point is reached at a temperature given by $dt(n)/dn= 0$, the maximum of $t(n)$. The fact that a maximum for $t(n)$ exists can be seen directly from eqn \ref{tn} as $t(n=0)$= $t(n=1/\alpha)$= 0 and therefore, a maximum must occur in between. At this maximum, $n(T)$ becomes undefined as adding or removing excitations does not change the temperature: $dn/dt \rightarrow \infty$ and we can no longer find a self-consistent solution of eqn \ref{tn}. We show this behavior in Fig. \ref{twolevel} where we plot $t(n)$ versus $n$ and 
$\Delta(n)$ versus $t(n)$.  One can see that there are two values for $\Delta(n)$ for each $T$, corresponding to a minimum and a maximum in the free energy; where the two solutions meet and terminate is the fold instability. We next discuss the 2-level system in terms of the free energy.\\
For a two-level system with excited-state occupation $n$,  we seek a free energy
$ F(n,T)=E(n)-TS(n)$ whose equilibrium condition
$\frac{\partial F}{\partial n}=0$ is consistent with eqn (\ref{occ}).
Using $\frac{dS}{dn}=k_B\ln\left(\frac{1-n}{n}\right)$, 
we find
\begin{equation}
\frac{dE}{dn}=\Delta(n);F(n,T)=\int_0^n \Delta(n')\,dn'-TS(n).
\label{freeenergy}
\end{equation}
Thus the self-consistent gap should be interpreted as the
incremental energy cost of increasing the excited-state
population, rather than as the energy per excitation.\\
For $\Delta(n)=\Delta_0\sqrt{1-\alpha^2 n^2}$,
we then obtain $F(n,T)$ which we differentiate with respect to $n$ to give
\begin{equation}
\frac{1}{\Delta_0}\frac{\partial F}{\partial n}
=\sqrt{1-\alpha^2 n^2}+t\ln\left(\frac{n}{1-n}\right).
\end{equation}
The equilibrium condition $\partial F/\partial n=0$ yields eqn (\ref{tn}).\\
The curvature of the free energy is
\begin{equation}
\frac{1}{\Delta_0}\frac{\partial^2 F}{\partial n^2}
=-\frac{\alpha^2 n}{\sqrt{1-\alpha^2 n^2}}
+\frac{t}{n(1-n)}.
\label{eq:Fnn}
\end{equation}
A stable equilibrium state requires $\frac{\partial^2F}{\partial n^2}>0$, so that the endpoint of the low-temperature branch is given by $\left.\frac{\partial^2F}{\partial n^2}\right|_{n_c,t_c}=0$.
Using $\partial F/\partial n=0$  to eliminate $t_c$, we find
\begin{equation}
1=\alpha^2 n_c^2\left[(1-n_c)\ln\left(\frac{1-n_c}{n_c}\right)\right],
\end{equation}
the same condition derived from
$\left.\frac{dT}{dn}\right|_{n_c}=0$.
Hence,
\begin{equation}
\frac{dT}{dn}=0
\quad\Longleftrightarrow\quad
\frac{\partial^2F}{\partial n^2}=0.
\label{equivalence}
\end{equation}
For $\alpha > 2$, this critical point is accompanied by a divergence of the specific heat. We see this by an expansion around the critical values $t_c$, $n_c$, and $\Delta_c$.  
The specific heat is $C=T\frac{dS}{dT}=T\frac{dS}{dn}\frac{dn}{dT}$ and  differentiating $F$ along
the equilibrium branch $F_n(n,T)=0$ yields
\begin{equation}
C=\frac{T k_B^2 \left[
\ln\left(\frac{1-n}{n}\right)
\right]^2}{F_{nn}}.
\label{Cfreeenergy}
\end{equation}
where subscripts denote partial derivatives.
Thus, the specific-heat singularity occurs because the curvature of the free-energy minimum vanishes, $F_{nn}\rightarrow0^+.$ Exactly at the critical point we have
$F_{nn}$= 0, so that the leading free energy terms in $T_c-T$ and $n_c-n$ are
\begin{equation}
\frac{\delta F}{\delta n}=0\simeq
F_{nT,c}(T-T_c)+\frac{1}{2}F_{nnn,c}(n-n_c)^2.
\label{triple}
\end{equation}
Therefore, $n-n_c\propto (T_c-T)^{1/2}$
and  $F_{nn}\propto n-n_c \propto|T_c-T|^{1/2}$. We see that the second derivative changes sign at $n=n_c$.  From this we obtain the singular behavior of the specific heat and the gap near $n_c$:
\begin{equation}
C(T)\propto(T_c-T)^{-1/2};
\frac{d\Delta}{dT}\propto -(T_c-T)^{-1/2}.
\label{Deltadivergence}
\end{equation}
Thus, while the gap remains finite at $T_c$, it approaches its critical
value with a vertical tangent (see Fig. \ref{twolevel}). When $F_{nn}$= 0, the specific heat diverges,  even in the absence of an order parameter that our two-level system is clearly lacking. 
The two-level model thus establishes that an occupancy-dependent damping rate can destabilize a gapped excitation spectrum without requiring the gap to close. We now examine whether the measured interactions between rotons in superfluid helium provide sufficient feedback to produce such an instability at the superfluid transition in $^4$He at $T_{\lambda}$ and  discuss its relevance to the hidden order transition (HO) in URu$_2$Si$_2$ at  $T_{HO}$. We argue that the thermodynamic stability threshold is breached in both systems at the transition temperature.\\
Liquid $^4$He undergoes a second order phase transition into a superfluid state at 2.172 K \cite{Kapitza1938}. This transition is marked by the characteristic $\lambda$-shaped curve of the specific heat, and below $T_{\lambda}$ the ground state supports a Bose condensate \cite{London1938, Hohenberg1965,Glyde2000} that furnishes the order parameter in the low temperature state. The excited states are separated in energy from the ground state and their dispersion \cite{Donnely1981} is referred to as the phonon-roton dispersion curve. The roton energy gap that occurs at a finite momentum is the defining feature when it comes to measurable properties such as the specific heat and entropy above 1 K \cite{Landau1947}. The excitations acquire a finite lifetime because of their interactions with the $n(T)$ other excitations that are present at finite temperatures $T$. This damping mechanism goes by the name of Landau-Khalatnikov (LK) damping \cite{Landau1949} and it has been shown that this mechanism accounts for the observed finite lifetimes of the excitations  below 1.5 K \cite{Andersen1996,Fak2012}.\\
Upon raising the temperature close to $T_{\lambda}$, the lifetimes rapidly decrease (see lower panel Fig. \ref{damping}) and the excitations cease to be elementary excitations, rather they transform in a continuous manner into the density fluctuations characteristic of a normal fluid \cite{Svensson1996}. Experimentally, this smooth transition has been scrutinized \cite{Svensson1996,Montfrooij2000} by means of neutron scattering experiments approaching $T_{\lambda}$ to within 0.001 K. Near and just above $T_{\lambda}$, the roton excitation becomes very poorly defined, decaying back to equilibrium before it has propagated by one wavelength. Concomitantly, the condensate fraction of the ground state is depleted \cite{Glyde2000} as expected since the phase coherence of the condensate cannot be maintained when the excited states are no longer clearly separated in energy from the ground state. Therefore, the $\lambda$-transition seems to be driven by a damping rate that becomes too large for the excitations to remain well defined, with the increase in damping rate attributable to the increasing population of excitations \cite{Montfrooij2010}. \\
As mentioned, the damping rate of the roton excitation at low temperatures ($T <$ 1.5 K) tracks the number of excitations according to LK-damping \cite{Andersen1996,Fak2012}. At these and  higher temperatures approaching $T_{\lambda}$, the roton spectra can be accurately modeled \cite{Svensson1996} by a damped harmonic oscillator (DHO) form. Above the transition, this description remains valid, revealing the strongly damped density fluctuations of the normal fluid phase. Based on this, we extend the  LK-damping rate [$\Gamma(T)=\Delta_0 \alpha n(T)$] to higher temperatures keeping $\alpha$ fixed \cite{rates}.  We use the standard expression for the number of rotons $N_{rot}(T)$ and evaluate $\Delta(T)$ using the DHO form to solve for the number of roton excitations per $^4$He atom $n_{rot}(T) \equiv N_{rot}(T)/n_0$ given by
\begin{equation}
n_{rot}(T)=\frac{k^2_{\text{rot}}\sqrt{2\pi \mu k_BT}}{2\pi^2 \hbar n_0}(1+b\sqrt{t})e^{-\Delta_0\sqrt{1-\alpha^2n^2_{rot}(T)}/k_BT},
\label{nrot}
\end{equation}
with $n_0$ the number density, $k_{\text{rot}}$ the roton wavenumber,  $\mu$ the effective mass for the parabolic roton excitation curve approximation and $b\sqrt{T}$ capturing the deviation from the parabolic form; all these values are known from experiment \cite{Fak2012}.
We show in Fig. \ref{lambda} that this approach qualitatively reproduces the very rapid $\Gamma(T)$-increase observed experimentally with the ensuing thermodynamic instability 0.3 K above  the actual transition temperature. We have also performed a more sophisticated calculation \cite{calcs} following the methodology in Ref. \cite{Fak2012} and found a virtually indistinguishable curve. These findings demonstrate that the $\Gamma \sim n$ term can produce an instability, but at a temperature that exceeds $T_{\lambda}$.\\
Inclusion of the energy shift term $\delta \epsilon = \Delta_0 \beta n_{\text{rot}}(T)$ (originating from the same 4-vertex interaction that determines the LK-damping mechanism as well as from the Hartree term) lowers the instability point close to the actual $T_{\lambda}$. We added this shift to eqn \ref{nrot} [replacing the square root with $\sqrt{(1+\beta n)^2-\alpha^2n^2}$] and again solved in a self-consistent manner using the published values \cite{Fak2012} for $\alpha$ and for the shift due to roton-phonon interaction as well as the shift due to roton-roton interactions. For the latter we reduced the published \cite{Fak2012} value by $\sim$ 30\% in order to  drive the instability threshold to 2.172 K, bringing $\beta$ close to an earlier measurement \cite{earlier,footnote}. The results are shown as the solid curve in Fig. \ref{lambda}. Note that only including the shift but not the damping \cite{Jacklenote} does not result in an instability in $^4$He consistent with experimental values. However, including the damping and benchmarking the unknowns $\alpha$ and $\beta$ to their measured low-T values \cite{Fak2012,earlier} yields a physically consistent description of the $\lambda$-transition, making it evident that the transition is occupancy driven with the LK-mechanism determining the transition temperature.\\
In Fig. \ref{lambda} we also display the experimental values \cite{Montfrooij1997}  for $\Gamma(T)$ which increase very rapidly on approaching $T_{\lambda}$. We note that at low temperatures, the density fluctuation created by a neutron couples predominantly to the well-defined roton which is clearly separated from the other many-body excitations. Approaching $T_{\lambda}$, however, the neutron-created disturbance increasingly overlaps with a broad superposition of many-body eigenstates. Consequently, the experimentally observed linewidth  pertaining to  the demise of the density fluctuation can substantially exceed the calculated damping  within LK-theory. This explains why the measured $\Gamma(T)$ (Fig. \ref{lambda}) exceeds the calculated values near the transition, however, both experiment and calculation show the very rapid increase that characterizes a fold instability. As a final check we display (Fig. \ref{lambda}) the point at which the roton density equals the fluid density \cite{Jackle1972}. We observe that this independent check coincides with the critical density at the fold instability.\\
Given the known $\Gamma(T)$-dependence at low $T$ and the accuracy of the DHO description on either side of the transition, combined with our results in Figs. \ref{damping} and \ref{lambda}, we conclude that the superfluid to normal fluid transition is driven by the number of excitations reaching a critical value $n_c$. The roton excitation ceases to be a thermodynamically stable excitation and the excited states become too intertwined with the ground state to maintain the phase coherence of the condensate, and the superfluid order parameter drops to zero. This view is entirely consistent with the fact that the entire phonon-roton curve can be numerically reproduced \cite{Jackson1962,Montfrooij2000} using perturbation theory with the initial estimates for the excited states given by propagating density fluctuations, independent of the condensate fraction of the ground state. Thus, in $^4$He, the occupancy drives the transition from superfluid to normal fluid.\\
URu$_2$Si$_2$ undergoes a phase transition at $T$= 17.5 K from a high temperature Kondo screened metallic phase to a low-T phase characterized by ground and excited states that are both (extended) magnetic singlets \cite{Broholm1991,Mydosh2011,Mydosh2020}. The excited state dispersion displays a momentum-dependent energy gap throughout the Brillouin zone \cite{Broholm1991}. At the transition, the specific heat exhibits a $\lambda$-shape, but 4 decades of research \cite{Mydosh2020} have failed to positively identify an order parameter for this    transition, explaining why this transition is referred to as the hidden order transition.\\
Recently, it was shown \cite{Montfrooij2024} that low temperature  phases of URu$_2$Si$_2$ and $^4$He can be mapped onto each other: scaling the temperature by the relevant energy gap, both the entropy and the damping rates of excitations (Fig. \ref{damping}) were shown to have an identical temperature evolution, strongly suggesting that the transition in URu$_2$Si$_2$ is also driven by thermal occupancy. While this clarified the mechanism and the source of the entropy shed at the hidden order transition, it did not identify an order parameter.\\
In detail, for $T< T_{HO}$, the ground state and excited states  in URu$_2$Si$_2$  are extended magnetic singlets separated by a gap, with an increasing number of excitations appearing upon raising the temperature. The high temperature phase ($T>T_{HO}$) is a Kondo shielded phase where the gapless excited state and ground state function as an effective Kondo doublet. At $T_{HO}$, the shielding is not complete \cite{Bretana2023} leaving $\sim$0.2 Rln2 in entropy in the system at $T_{HO}$ \cite{Montfrooij2024}. This is the exact amount that is released when the system is cooled through $T_{HO}$ \cite{Mydosh2011}. While considerable effort has been devoted to identifying an order parameter for the low-$T$ phase, there is actually no particular reason why a phase in which the two lowest levels are magnetic singlets should have an order parameter associated with it; in fact, it would be unexpected. Given that the phase transition is almost certainly driven by thermal occupancy (Fig. \ref{damping}) which already would account for the specific heat divergence, combined with Kondo shielding being incomplete \cite{Bretana2023,Montfrooij2024} so that residual entropy is available on going from one  phase to another, we consider it very likely that the hidden-order transition can be understood as an occupancy-driven instability without having to invoke a symmetry-breaking order parameter. \\
We expect occupancy-driven instabilities to be widespread in both gapped and gapless systems, whether the feedback arises from energy renormalization through the real part of the self-energy $\Sigma$ or from damping through its imaginary part. In the spirit of Eq.~\ref{nrot}, whenever the self-consistency condition can be written as $n=H(n,T)$ (with $H$ containing all possible feedback mechanisms), a generic fold instability occurs when $\partial H/\partial n=1$. We note that such an instability need not coincide with an equilibrium phase transition, since a competing phase may become thermodynamically favorable before the self-consistent solution disappears. Nevertheless, while damping is conventionally viewed as a consequence of approaching a phase transition, our results demonstrate that feedback between thermal excitation populations on the one hand and both damping and energy shift on the other hand can itself destabilize a phase.


\begin{thebibliography}{99}
\bibitem{Jacklenote} J\"ackle already noted \cite{Jackle1972} that damping would be required for a better agreement, but did not view damping as the essential factor in creating the instability.
\end{thebibliography}
\end{document}